\documentclass{elsarticle}

\usepackage{lineno,hyperref}
\usepackage{amsmath}
\modulolinenumbers[5]

\journal{Icarus}

\biboptions{authoryear}

\begin{document}

\begin{frontmatter}

\title{Tropical and Extratropical General Circulation with a Meridional Reversed Temperature Gradient as Expected in a High Obliquity Planet}



\author[harvardseas]{Wanying Kang \corref{mycorrespondingauthor}}
\cortext[mycorrespondingauthor]{Wanying Kang}
\ead[url]{wanyingkang@g.harvard.edu}

\author[fsu]{Ming Cai}

\author[harvardseas,harvardeps]{Eli Tziperman}

\address[harvardseas]{School of Engineering and Applied Sciences, Harvard University, 20 Oxford st., Cambridge, Massachusetts, USA}
\address[harvardeps]{Department of Earth and Planetary Science, Harvard University, 20 Oxford St., Cambridge, Massachusetts, USA}
\address[fsu]{Department of Earth, Ocean and Atmospheric Sciences, Florida State University, 1017 Academic Way, Tallahassee, FL, USA}

\begin{abstract}  
Planets with high obliquity receive more radiation in the polar regions than at low latitudes, and thus, assuming an ocean-covered surface with sufficiently high heat capacity, their meridional temperature gradient was shown to be reversed for the entire year. The objective of this work is to investigate the drastically different general circulation of such planets, with an emphasis on the tropical Hadley circulation and the mid-latitude baroclinic eddy structure.

We use a 3D dry dynamic core model, accompanied by an eddy-free configuration and a generalized 2D Eady model. When the meridional temperature gradient is reversed, the Hadley cell becomes much weaker, shallower and thermally indirect, as seen by other studies, though not in the context of high obliquity planets. For both the normal and reverse temperature gradient cases, we show that the surface friction and eddy momentum transport are the primary drivers of the Hadley cell. Because mid-latitude baroclinic eddies concentrate westerly momentum toward the mid-latitude baroclinic zone, the surface wind pattern is easterly-westerly-easterly from the equator to the pole in both cases. As a result, low-latitude air parcels near the surface gain westerly momentum through friction, and are redirected equatorward by the Coriolis force, forming thermally indirect (direct) circulation in the reverse (normal) case. The Hadley cell under a reverse temperature gradient configuration is shallow and weak, even when the magnitude of the gradient is the same as in the normal case. This shallow structure is a result of the bottom-heavy structure of the baroclinic eddies in the reverse case, and the relatively weak wave activity. We propose a new mechanism to explain the mid-latitude eddy structure for both cases, and verify it using the generalized Eady model. With seasonal variations included, the annual mean circulation resembles that under perpetual annual mean setup. Approaching the solstices, a strong cross-equator Hadley cell forms in both cases, and about 2/3 of the Hadley circulation is driven by eddies, as shown by eddy-free simulations and using a decomposition of the Hadley cell.
 \end{abstract}


\end{frontmatter}


\section{Introduction}

Exoplanets, including the habitable ones, may have a large obliquity or large obliquity variance, depending on the initial angular momentum of the nebulae that form the planet, continental movement \cite[][]{Williams-Kasting-Frakes-1998:low}, gravitational interaction with other bodies \cite[][]{Correia-Laskar-2010:tidal}, and the history of orbital migration \cite[][]{Brunini-2006:origin}. In our solar system, for example, Mars's obliquity chaotically varies from 0 to 60 degree \cite[][]{Laskar-Robutel-1993:chaotic}, and Venus and Uranus have obliquities close to 180 and 90 degree respectively \cite[][]{Carpenter-1966:study}. Even among Hot Jupiters, usually close to the star, several planets were observed with high obliquity, in about 50 star systems \cite[][]{Winn-Fabrycky-Albrecht-et-al-2010:hot, Albrecht-Winn-Johnson-et-al-2012:obliquities}. 

Earth may also have been in a high obliquity state backward in time. The early Earth was sufficiently warm to allow life, in spite of the sun being 20-30\% weaker than at present \cite[][]{Sagan-Mullen-1972:earth}. High obliquity may help resolve this so-called ``Faint-Young Sun paradox'' without requiring an extremely high concentration of green house gases \cite[][]{Kasting-1987:theoretical, Kuhn-Walker-Marshall-1989:effect, Pavlov-Kasting-Brown-et-al-2000:greenhouse}. GCM simulations showed that only 12 times present level CO$_2$ is required to keep the global mean surface temperature above 279K, with 70$^\circ$ obliquity, compared to several bars of CO$_2$ suggested for warm climate \cite[][]{Jenkins-2001:high, Donnadieu-Ramstein-Fluteau-et-al-2002:is}. The high obliquity for the early Earth might result from a giant impact around 4.5 Ga, associated with the lunar formation \cite[][]{Williams-2008:proterozoic}. In addition, a high obliquity Earth may also explain the low latitude glacial events in the Early Proterozoic \cite[2.5-2 Gyr][]{Evans-Beukes-Kirschvink-1997:low, Schmidt-Williams-1995:neoproterozoic} and Late Proterozoic \cite[900-570 Myr][]{Park-1997:paleomagnetic}. With open water in the high latitudes as oppose to the ``snowball Earth'' hypothesis \cite[][]{Kirschvink-1992:late, Hoffman-Kaufman-Halverson-et-al-1998:neoproterozoic}, the high-obliquity hypothesis avoids the difficulty to explain the triggering and termination of the ``snowball Earth'' state \cite[][]{Pierrehumbert-2005:climate, Hu-Yang-Ding-et-al-2011:model}. Also, recent observation studies identified strong seasonality near the paleoequator based on cryogenian deposits from many sites, again accommodated better by the high obliquity hypothesis than by the ``snowball Earth'' or ``slushball Earth'' hypothesis \cite[][]{Williams-2008:proterozoic, Williams-Schmidt-Young-2016:strongly}. Thus, a high obliquity configuration may also be relevant for Earth paleoclimate. 

High obliquity climate has been investigated by many studies, using energy balance models \cite[][]{Williams-Kasting-1997:habitable, Gaidos-2004:seasonality, Spiegel-Menou-Scharf-2009:habitable}, and using 3D general circulation models \cite[][]{Jenkins-2001:high, Donnadieu-Ramstein-Fluteau-et-al-2002:is, Williams-Pollard-2003:extraordinary, Ferreira-Marshall-OGorman-et-al-2014:climate, Linsenmeier-Pascale-Lucarini-2015:climate, Wang-Liu-Tian-et-al-2016:effects, Kilic-Raible-Stocker-2017:multiple}. Paleoclimate studies by \cite{Jenkins-2001:high, Donnadieu-Ramstein-Fluteau-et-al-2002:is}, and \cite{Williams-Pollard-2003:extraordinary} reported habitability at high obliquity despite a much stronger season cycle, and \cite{Linsenmeier-Pascale-Lucarini-2015:climate} noticed a substantial outward expansion of the habitable zone under high obliquity states. Given the importance of the atmospheric general circulation in determining the surface temperature distribution and thus the habitability, studies on understanding the atmospheric general circulation are needed.

In this work, we look into the general circulation of an atmosphere under a reversed meridional temperature gradient (reverse case), as expected in a high obliquity planet with high enough surface heat capacity \cite[][]{Williams-Holloway-1982:range, Jenkins-2001:high, Ferreira-Marshall-OGorman-et-al-2014:climate}. The Hadley cell driven by the reverse temperature gradient does not reverse as well, and instead is found to be thermally indirect, and to be much weaker and more bottom-amplified compared to the normal case driven by the same magnitude of the non-reversed temperature gradient. This thermally-indirect Hadley cell is shown to be driven by friction, which has to counterbalance the column-integrated eddy momentum drag. It is also shown that the bottom heaviness stems from the bottom-amplified structure of the most unstable baroclinic normal mode. A High obliquity planet is expected to experience a strong seasonal cycle, and we also consider the circulation in this case. We find that baroclinic eddies during the solstices only prevail in one hemisphere, transporting heat downgradient and driving a strong cross-equatorial Hadley cell. We also find that the annual mean of the seasonal general circulation is similar to the circulation calculated under perpetual annual mean forcing.

  We use a dry-core atmospheric model with a restoring to a prescribed radiative equilibrium temperature. We contrast two scenarios: one with normal negative meridional temperature gradient and the other with a reversed gradient, which allows us to isolate the effects of the direction of the temperature gradient without concerning the impact of obliquity variation on the amplitude of meridional temperature gradient. These equilibrium temperature profiles are chosen to allow a clean comparison between inverse and normal gradients, as using a representation of the high and low obliquity radiation may lead to changes of both amplitude and sign of the temperature gradient.

Idealized experiments with a reversed meridional temperature gradient has been used by \citet[][hereafter AS15]{Ait-Chaalal-Schneider-2015:why}, to study why, on the Earth, the eddy meridional momentum transport concentrates in the upper troposphere. The resulting general circulation was shown to be entirely different from the reversal of the Earth-like circulation, featuring a thermally-indirect Hadley cell in the tropics, a thermally-direct Ferrel cell in the mid-latitudes, and eddy momentum transport concentrated near the surface, all consistent with ours.

In section~\ref{sec:Hadley-vanishment}, we present and examine three possible mechanisms to explain the drastic different Hadley cells in the normal and reverse cases. The bottom amplified eddy momentum transport was argued by AS15 to be related to the vertical Rossby wave propagation in the troposphere. In section~\ref{sec:baroclinic-mode}, we propose an alternative mechanism associated with the vertical structure of the most unstable baroclinic mode, and support it using the modified Eady model. The baroclinic eddy features are then used to explain the general circulation features including the Ferrel cell, Hadley cell and the surface wind. We finally add an idealized seasonal cycle (section~\ref{sec:seasonal-cycle}), and find the solstice Hadley cell to be also largely driven by midlatitude baroclinic eddies, but through their heat transport rather than momentum transport, as we find in the perpetual annual mean configuration. Given the fact that the baroclinic eddies are weaker and shallower in the reverse temperature gradient case, the Hadley cell is also weaker and shallower. 


\section{Model configurations}
\label{sec:methods}

\subsection{Held-Suarez type model}
\label{sec:HS-model}
We use a dry dynamic core model with highly simplified physics \cite[][]{Held-Suarez-1994:proposal} to investigate the general circulation in a reversed temperature gradient climate. Two idealized experiments, one with normal negative meridional temperature gradient and one reversed, are set up based on the idealized physics component set in the Community Earth System Model version 1.2.2 \cite[CESM,][]{Neale-Chen-Gettelman-2010:description}, where the radiation, convection and other physics processes are parameterized by a relaxation to the radiative equilibrium temperature profile, $T_{eq}(\phi,p)$. All settings for the normal case are as in \cite{Held-Suarez-1994:proposal}. To represent the surface heating and friction, below 700 mb toward the ground, the temperature relaxation time scale gradually decreases from 40 days in the free atmosphere to 4 days, plus additional linear wind damping whose timescale linearly decreases to 1 day.  For the reverse case, we revert the $T_{eq}$ between the equator and the poles to force a reversed meridional temperature gradient. We write the $T_{eq}$ for both cases in the following generalized form,
\begin{eqnarray}
  \Theta_{eq}&=&\max\left\{200 K, T_c + T_y \left(\sin\phi -\sin\phi_c\right)^2 + S\ln{(p_0/p)} \right\}\nonumber\\
  T_{eq}&=&\Theta_{eq}\left(\frac{p}{p_0}\right)^\kappa, \label{eq:Teq}
\end{eqnarray}
where $\phi_c$ is the center latitude, where the maximum (minimum) surface temperature $T_c$ is achieved in the normal (reverse) case. The center latitude is set to $0$ to represent the equinox except in section~\ref{sec:seasonal-cycle}, where a seasonal cycle is added. $p_0=100000$ Pa is the reference pressure, and $\kappa=2/7$. $p$ denotes the pressure, and $\phi$ denotes the latitude. $T_y$ is the equator to pole temperature difference. 
With $T_c=315$ K, $T_y=-60$ K, $\phi_c=0$, $S=10\cos^2\phi$ K, we recover the profile used by \cite{Held-Suarez-1994:proposal}; while in the reverse case, the settings, $T_c=255$ K, $T_y=60$ K, $\phi_c=0$, $S=10\sin^2\phi$ K, are used to produce the mirror reflection of the original profile about 45 degree latitude for each hemisphere.

We use the finite volume dynamic core, with a horizontal resolution of 1.875$^\circ$ in latitude and 2.5$^\circ$ in longitude. There are 35 vertical levels from the surface to the model top at 3 mb, of which 18 layers are in the stratosphere, which guarantees the vertical resolution is finer than 2 km throughout the domain.

This idealized dry core model is appropriate for studying the dynamic asymmetry between poleward and equatorward heat transports in the normal or inverse temperature gradient cases. However, given the highly simplified physics in our model and the idealized equilibrium temperature profiles being used, the model does not mean to fully represent the realistic circulation of either scenario.

\subsection{Decompose meridional circulation.}
\label{sec:decompose-meridional-circulation}
Following a similar path in \cite{Kuo-1956:forced}, we evaluate the contributions by eddies, diabatic heating, and friction, by solving the meridional streamfunction forced by each component alone. Start from the temperature and zonal momentum equation,

\begin{align}
  \label{eq:T-U-equation}
  &\partial_t U-fV = -\frac{V}{a\cos\theta}\partial_\theta (U\cos\theta) -\Omega \partial_p U-\frac{1}{a\cos^2\theta}\partial_\theta(\overline{u'v'}\cos^2\theta)-\partial_p(\overline{u'\omega'})+F\\
&\partial_t T - S_p\Omega = -\frac{V}{a}\partial_\theta T -\frac{1}{a\cos\theta}\partial_\theta(\overline{v'T'}\cos\theta)-\gamma\partial_p(\overline{\omega'\theta'})+\frac{Q}{C_p},
\end{align}
where $F$ is friction, $Q$ is diabatic heating, $S_p=-\gamma\frac{\partial \overline{\Theta}}{\partial p}$ is the static stability, and $\gamma=(p/p_0)^\kappa$. $\overline{(\cdot)}$ denotes time mean zonal mean, and prime denotes deviations. Cancel the time derivative of $T$ and $U$ in both equations using thermal wind balance, $pf U_p= R/a T_\theta$, yields


\begin{align}
&fpM_y\frac{\partial^2\psi}{\partial p^2}+\frac{RS_p}{a^2}\cos\theta\frac{\partial}{\partial\theta}\left(\frac{1}{\cos\theta}\frac{\partial\psi}{\partial\theta}\right)-\frac{f(1-\kappa)}{a}\frac{\partial U}{\partial p}\frac{\partial \psi}{\partial\theta}+\frac{pf\cot\theta}{a^2}\frac{\partial U}{\partial p}\frac{\partial\psi}{\partial p}=\nonumber\\
&+\frac{fp}{a\cos\theta}\frac{\partial^2\left(\overline{u'v'}\cos^2\theta\right)}{\partial\theta\partial p}+fp\cos\theta\frac{\partial^2\overline{u'\omega'}}{\partial p^2}-\frac{\cos\theta}{a^2}\frac{\partial}{\partial_\theta}\left(\frac{1}{\cos\theta}\frac{\partial\left(\overline{v'T'}\cos\theta\right)}{\partial\theta}\right) -\frac{\gamma R\cos\theta}{a}\frac{\partial^2\left(\overline{\omega'\theta'}\right)}{\partial\theta\partial p}\nonumber\\
  &-fp\cos\theta\frac{\partial F}{\partial p}+\frac{R}{C_p}\frac{\cos\theta}{a}\frac{\partial Q}{\partial \theta}  \label{eq:hadley-drive}\\
  &M_y=f-\frac{1}{a\cos\theta}\frac{\partial}{\partial\theta}(U\cos\theta)
\end{align}

where $R$ is gas constant, $a$ is the earth radius, $\psi$ is the meridional streamfunction defined as below:
\begin{align}
  &V=\frac{1}{\cos\theta}\frac{\partial\psi}{\partial p}\\
  &\Omega=-\frac{1}{a\cos\theta}\frac{\partial \psi}{\partial \theta}.
\end{align}

Eq~\eqref{eq:hadley-drive} will be an elliptic equation as long as the meridional temperature gradient (or equivalently the vertical wind shear) is not too strong. By solving Eq.~\eqref{eq:hadley-drive}, forced by each term on the right hand side, we get the meridional circulation driven by each forcing.

There are two purposes to do such cancellation: First, the vanishment of the time derivative terms allow us to decompose the meridional circulation for one particular season in the seasonal varying experiments (section~\ref{sec:seasonal-cycle}) without worrying about the climatology change through that season. Second, the advection of geostrophic $U$ and $T$ will also be largely canceled, left the left hand side (LHS) terms of Eq~\eqref{eq:hadley-drive}. Some of these LHS terms are $U$ related. We have tested that even if we move those $U$-related terms to the right hand side as an external forcing, and solve the rest poisson-like equation,
\begin{align}
  \label{eq:poisson}
  f^2p\frac{\partial^2\psi}{\partial p^2}+\frac{RS_p}{a^2}\cos\theta\frac{\partial}{\partial\theta}\left(\frac{1}{\cos\theta}\frac{\partial\psi}{\partial\theta}\right)=\ldots,
\end{align}
the result of meridional circulation decomposition remains almost the same. The physics picture is more clear in Eq~\eqref{eq:poisson}. Forced by one single forcing, the distribution of $U$ and $T$ could be different from those forced by all forcings, however, as long as the change follows the thermal wind relationship (which should be the case), we can evaluate the meridional circulation being driven without knowing $U$ and $T$. In the results section, we stick to Eq~\eqref{eq:poisson}.

Total meridional circulation is evaluated by taking vertical integral of meridional mass transport.

\begin{align}
  \label{eq:total-psi}
  \psi=\frac{2\pi a\cos\theta}{g}\int^p_0~V~dp'.
\end{align}
To match the unit, we multiply a factor, $\frac{2\pi a}{g}$ to all forced streamfunction terms solved from the poisson equation.

\subsection{Generalized Eady model}
\label{sec:Eady-model}
To understand the different baroclinic wave behaviors in the normal and reverse cases, we generalized the original Eady model \cite[][]{Eady-1949:long} to allow a vertically varying profile of zonal wind and static stability as well as
to include the beta effect (which is set to zero in the original Eady problem). The governing PDEs are the QGPV conservation law in the interior plus rigid boundary condition at the top and bottom,

\begin{eqnarray}
  \left[\frac{\partial}{\partial t} + ik U(z)\right] q' + ik Q_y\psi'&=&0, \hspace{1cm}0<z<H \\
  \left[\frac{\partial}{\partial t} + ik U(0)\right] \psi'_z - ikU_z \psi'&=&0, \hspace{1cm} z=0\\
  \left[\frac{\partial}{\partial t} + ik U(H)\right] \psi'_z - ikU_z \psi'&=&0, \hspace{1cm} z=H
\end{eqnarray}

\begin{eqnarray}
  q'&=& -k^2\psi'+\frac{f^2}{N^2}\psi'_{zz} - \frac{f^2}{N^2}\left(\frac{1}{H_\rho} + \frac{\partial \ln{\left(N^2\right)}}{\partial z}\right)\psi'_z
\end{eqnarray}
In above equations, $z=H\log_{10}(p_0/p)$, where $p_0=10^5$ Pa and $H=10$ km. $H_\rho$ is the density e-fold scale which is set to $H/\ln{10}$ ($H$ and $H_\rho=\frac{RT}{g}$ are constant, and that is where the isothermal approximation comes in), $N^2$ is the squared Brunt Vasala frequency. $k$ is the zonal wavenumbers of the perturbations and is chosen to be $k=5$.
Our generalized Eady model intends to represent the longitude-vertical slice at 45 N for the normal case and at 55N for the reverse case. We made the choice by looking for the latitudes where baroclinic eddy heat transport peaks in the Held-Suarez GCM simulations. The Coriolis coefficients, $f$ and $\beta$, and the perimeter of the zonal circle are set to represent the corresponding latitudes. The model has 101 layers, uniformly distributed in $\ln(p)$ coordinate, representing the bottom 15 km of atmosphere, including troposphere and the bottom stratosphere. The stratosphere is necessary to damp the wave before it encounters the top. The vertical profiles of zonal mean zonal wind $U(z)$, Brunt Vasala frequency $N(z)$, and meridional QGPV gradient $Q_y(z)$ are specified in the supplementary material, and are compared with the Held-Suarez model output (Fig.~S1). To identify the mode structure without the exponential growth, we normalize the perturbation streamfunction to have unit norm, at each time step.

\section{Results}
\label{sec:results}

Since the meridional gradient of the radiative equilibrium temperature in the reverse temperature gradient case (hereafter reverse case) is equal and opposite to that in the case with temperature decreasing with latitude (hereafter normal case), one might anticipate that the solution would simply reverse as well; that includes a reversed Hadley cell, Ferrel cell, surface wind pattern, heat transport and etc. This turns out to be not true, as there are other factors breaking the symmetry.

We start from the perpetual annual mean forcing in section~\ref{sec:Hadley-vanishment}, showing explicitly that the Hadley cell in the reversed case is eddy-driven, as is already known for the normal case \cite[][]{Bordoni-Schneider-2010:regime}. Next (section~\ref{sec:baroclinic-mode}), we study the vertical structure of the eddies, given their key role in driving the shallow reverse case Hadley cell. We propose a mechanism for the distinct eddy structures in the normal and reverse cases, alternative to that proposed by AS15, and discuss the implied general circulation features. Finally, we consider the seasonal cycle in a high obliquity climate (section~\ref{sec:seasonal-cycle}). The annual mean meridional circulation resembles that under a perpetual annual mean setup. The solstice baroclinic eddy structure remains similar but the eddy activity is strong in only one hemisphere. The heat transport by these eddies drives a strong Hadley cell crossing the equator.

\subsection{Hadley cell vanishes under the reverse temperature gradient setup}
\label{sec:Hadley-vanishment}
\begin{figure*}[!tbh]
 \centering
 \includegraphics[page=1,width=0.6\textwidth]{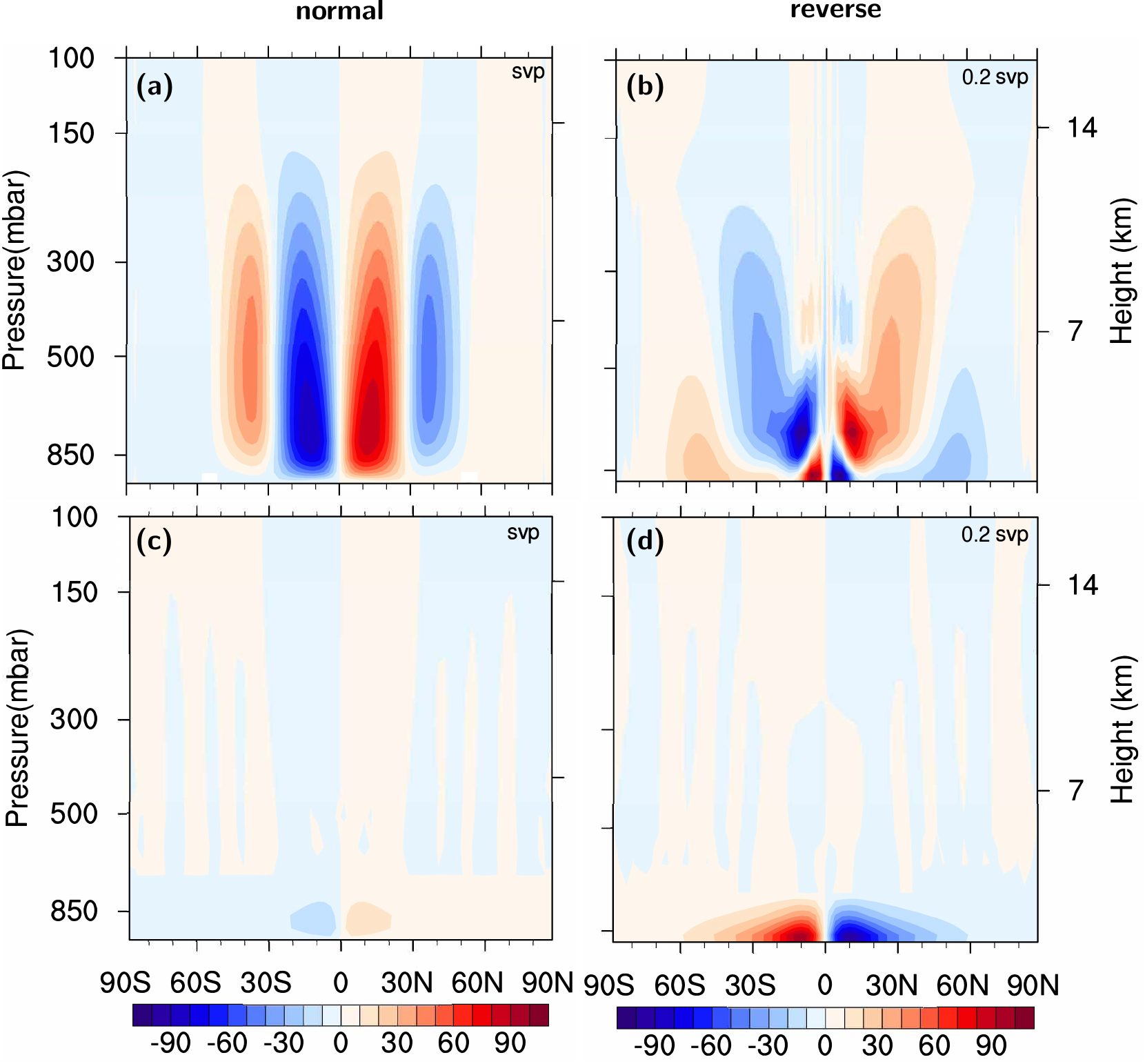}
 \caption{Zonal mean meridional circulation streamfunction for the normal case (a,c) and reverse case (b,d). (a,b) are for full eddy simulations and (c,d) are for eddy-free simulations. The streamfunction for the reverse cases (b,d) are in unit 0.2 svp rather than 1 svp as in the normal case. The maxima values are 89 svp, 20 svp, 19 svp and 19 svp respectively for the panel (a,b,c,d).}
 \label{fig:meridional-circulation-tropo}
\end{figure*}

Shown in Fig.~\ref{fig:meridional-circulation-tropo}(a,b) are the zonal mean meridional circulation streamfunction for the normal and reverse cases. In response to the reversed meridional temperature gradient, the reverse case Hadley cell does not change its direction, meaning that it becomes thermally indirect. In the meanwhile, the Hadley cell magnitude is 5 times weaker, with the total transport (the maximum of the meridional streamfunction) decreasing from 89 Sverdrup (hereafter svp) in the normal case to 20 svp. The Hadley cell structure also changes to be more concentrated toward the equator and the lower troposphere, compared to the normal case. All these results are consistent with AS15, where a reversed temperature gradient case is used to investigate why the Earth's eddy momentum flux is concentrated in the upper atmosphere.

We explore here three possible mechanisms that may explain why the Hadley cell in the reversed case does not resemble a reversed version of the normal Hadley cell, and show that only the first explanation is relevant. 1) Although the meridional temperature gradient is reversed, the mid-latitude eddy momentum transport remains in the same direction, concentrating westerly momentum toward the mid-latitude wave source region \cite[][]{Vallis-2006:atmospheric}. If the equinox Hadley cell is largely driven by eddies as it is in the Earth \cite[][]{Walker-Schneider-2006:eddy}, the Hadley cell direction in the reverse case would be expected to be the same as in the normal case, ascending in the deep tropics and descending in the subtropics. This circulation is thermally direct for the normal case, but thermally indirect for the reverse case. 2) As we reverse the meridional temperature gradient, the meridional distribution of the equilibrium stratification is also reversed, resulting in weaker tropical stratification, which might enhance the Hadley cell in the reverse case. 3) Even if the Hadley cell is driven by the meridional thermal contrast rather than by the eddies, there is still an asymmetry between the normal and reverse cases. That is, in the normal case, the westerly thermal wind above the equator violates the Hide's theorem -- there is no angular momentum extrema in the interior, and thus will drive a meridional overturning circulation \cite[][]{Emanuel-2007:quasi}. But in the reverse case, Hide's theorem is not violated, and thus the meridional overturning circulation may not need to form.

That the Hadley cell is weaker in the reverse case has already excluded mechanism (2), since a weaker stratification leads to a stronger Hadley cell rather than a weaker one. To completely exclude this mechanism, we set the static stability coefficient in Eq\eqref{eq:Teq} to $S=5$ K everywhere for both cases, and found the meridional circulation (not shown) to be almost identical to Fig.~\ref{fig:meridional-circulation-tropo}(a,b).  

Whether the eddies play an important role as proposed in mechanism (1) is examined through eddy-free experiments, where we set all wave components to zero (i.e., $k\geq 1$), following \cite{Becker-Schmitz-Geprags-1997:feedback} and \cite{Bordoni-Schneider-2010:regime}. The zonal mean meridional streamfunction of the corresponding eddy-free cases are shown in Figs.~\ref{fig:meridional-circulation-tropo}(c,d). Without eddies, the meridional overturning circulation is purely thermally-driven; this component is almost equal and opposite in the normal and reverse cases (notice that the colorbars are different in panels c and d), as expected from the same meridional temperature gradient magnitude in the two cases. Instead of being constrained in the tropics, the eddy-free Hadley cells expands all the way to the poles as the Hadley is the only way to transport heat, although they are very shallow. Compared to the Hadley cell in the full-eddy normal case, the eddy-free counterpart is much weaker, narrower and shallower, suggesting the normal case Hadley cell is mainly driven by eddies in the equinox setup, consistent with \cite{Becker-Schmitz-Geprags-1997:feedback} and \cite{Bordoni-Schneider-2010:regime}.
As for the reverse case, the thermally indirect Hadley cell seen in Fig.~\ref{fig:meridional-circulation-tropo}b disappears in the eddy-free setup, suggesting this thermally indirect Hadley to be driven by eddies as well.

That the Hadley cells in the two eddy-free experiments are almost equal and opposite indicates that the mechanism (3) is irrelevant. It makes little difference to the Hadley cell whether the thermal wind close to the equator is eastward (normal case) or westward (reverse case), let alone the fact that this purely thermally driven meridional circulation is not the dominant component of the Hadley cell in the full-eddy simulation.

In addition to turning off the eddies, we evaluate the contributions by eddies, diabatic heating, and friction, by solving the meridional streamfunction forced by each component alone (see section.~\ref{sec:decompose-meridional-circulation} for details). Dominant terms, those that are associated with $\overline{u'v'}$, surface friction, and $\overline{u'omega'}$, are shown in Fig.~\ref{fig:decomposition-perpetual-annual}. Also shown is the meridional streamfunction calculated by taking vertical integral of meridional velocity (Eq.~\ref{eq:total-psi}).

In both cases, these three terms, particularly the first two, explain the most parts of the total meridional circulation, meaning that the circulation is mostly momentum driven. The role played by vertical eddy momentum transport, $\overline{u'\omega'}$ is minor in the normal case, while it plays a significant role in the reverse case.
In section~\ref{sec:baroclinic-mode}, we will show that the eddy meridional momentum transport peaks in the middle troposphere in the normal case, while it is constrained near the surface in the reverse case. They cause the tropical air parcel lose westerly momentum, in the middle troposphere and near the surface respectively, driving meridional circulation in opposite directions. However, surface friction has to counterbalance the column-integrated meridional momentum convergence, whose latitudinal profile should be negative-positive-negative from the equator to the poles, no matter which side is warmer. That is because the Rossby wave concentrate momentum from the wave sink to the wave source located in the mid-latitudes. Thus surface friction will drive meridional circulation in the same direction for both cases. The thermally indirect circulation in the reverse case is driven by friction to a large extent, and the rest is driven by downward momentum transport by baroclinic eddies. The weakness of reverse case meridional circulation stems from the weakness of baroclinic eddies. We refer the reader to section~\ref{sec:baroclinic-mode} for more discussions on the baroclinic eddy structure.

Fig.~\ref{fig:residual-circulation-tropo} shows the residual circulation (Eulerian circulation plus eddy-driven circulation). Both cases have a single-cell thermally direct circulation in most of the troposphere. In normal case, the Eulerian circulation contributes most of the residual circulation in the tropics, leaving the extratropics to be dominated by the eddy-driven circulation. In the reverse case, however, eddy-driven circulation dominates almost everywhere, even in the deep tropics, suggesting a dominant role of eddies in meridional heat transport. Above the main cell, the reverse case has another, weaker, and opposite cell. This cell is also thermally direct, given that equator-to-pole temperature gradient is negative in the stratosphere (not shown).

The Hadley cell in the normal gradient case is only about the half the strength as in observation. \cite{Becker-Schmitz-Geprags-1997:feedback} showed that, in the absence of the latent heating release in the tropical upper atmosphere, the pressure gradient in the upper troposphere cannot be established, giving rise to a weak and shallow Hadley cell. As an extra heating was added over the tropics, the Hadley cell was found to enhance dramatically in both eddy-permitting and eddy-forbidden simulations. \cite{Ferreira-Marshall-OGorman-et-al-2014:climate} found that on high obliquity planets, convective precipitation was constrained to the high latitudes, while more gentle large-scale precipitation occurs more in the tropics. The exclusion of latent heat release in the HS model may thus underestimate the upper-level temperature gradient in the tropics, and hence, of the thermally direct Hadley component. If the mid-latitude wave activity does not strengthen proportionally to the upper level temperature gradient, in the presence of water vapor, the thermally indirect eddy-driven component of Hadley cell could be dominated by the thermally direct component.

\subsection{Baroclinic eddies in the mid-latitudes}
\label{sec:baroclinic-mode}

The atmospheric energy transport in the tropics is accomplished by the Hadley cell as discussed in the previous section; while in the mid-latitudes, the transport is completed through baroclinic eddies, instead. In this section, we investigate the generation, vertical structure and vertical extent of the mid-latitude baroclinic eddies, using the HS model aided with a generalized Eady model, and briefly discuss the surface wind pattern and the Ferrel cell structure. This analysis helps understand the general circulation features, including that the reverse case Hadley cell is weak, and more concentrated in lower troposphere.

Shown in Fig.~\ref{fig:eddy-momentum-transport} and Fig.~\ref{fig:eddy-heat-transport} top panels are the meridional eddy momentum transport (MEMT), $\overline{u'v'}$, and the meridional eddy heat transport, $\overline{v'T'}$, in the two cases, respectively. By just reversing the direction of temperature gradient, the structures of the baroclinic eddy transports change dramatically.
The meridional eddy momentum transport is poleward in both cases, thus driving a poleward motion in the tropical free troposphere (upper branch of the Hadley cell). But instead of occupying the whole mid-troposphere as it is in the normal gradient case, the reverse case eddy momentum transport is constrained close to the surface and is much weaker, accompanied with another even weaker peak in the middle to upper troposphere. The meridional eddy heat transport in the normal case has two maximums, one near 200 mb and the other near the surface; while, in the reverse gradient case, there is only one center close to the surface and it is again much weaker.

We extract the baroclinic eddy structures from the HS experiments, using the method in \cite{Lim-Wallace-1991:structure}. We project the 3D daily geopotential height field onto the standardized geopotential time series at (180E, 45N, 700 mb). Shown in the upper panels of Fig.~\ref{fig:eddy-structure} are the wavenumber 7 eddy structure diagnosed from HS experiments. For the normal (reverse) case, the baroclinic eddy structures tilt westward (eastward) vertically, allowing poleward (equatorward) heat transport. Also, consistently, there are two amplitude peaks for both cases, but the upper one is stronger in the normal case, and the lower one is stronger in the reverse case. These eddy structures are not sensitive to the longitude and pressure level choice of the projected time series, and are valid in $\pm 10$ deg change in latitude.

The shift of MEMT peak from the upper troposphere to the surface with the reversal of meridional temperature gradient has also been noticed by AS15. The authors falsified two previous hypotheses on the MEMT structure under the Earth-like configuration, respectively regarding the surface friction damping effect \cite[][]{Held-2000:general, Vallis-2006:atmospheric}, and the more conducive wave propagation environment in the upper troposphere \cite{Held-2000:general, Held-2007:progress}; and they raised up the following hypothesis. To extract energy from the baroclinicity, the meridional eddy heat transport (proportional to the vertical component of wave activity flux) was negative in the reversed temperature gradient case; as waves propagate downward, they encountered the surface, and turned meridionally there (note that the meridional component of wave activity flux is proportional to the meridional momentum transport). In the Earth-like case, waves propagate upward and then turn meridionally near the tropopause instead. 


We propose an alternative mechanism to explain the baroclinic wave vertical structure: with the opposite meridional temperature gradient (vertical wind shear), the most unstable baroclinic mode structure changes from westward-tilting top-amplified to eastward-tilting bottom-amplified, and the eddy momentum and heat transports also change accordingly.

To test the hypothesis, we generalize the single wavenumber Eady model, to count in realistic vertical profiles of zonal wind, meridional QGPV gradient and stratification. The analytical formula of these profiles are given in supplementary material, and they are compared against the HS model in Fig.~S1. We choose to represent the vertical slice at 45N and 55N latitude in normal and reverse cases, because these latitudes are roughly where westerly momentum is converged to by eddies (see Fig.~\ref{fig:eddy-momentum-transport}) and where eddy heat transport (see Fig.~\ref{fig:eddy-heat-transport}) peaks.

$k=5$ is chosen for both normal and reverse gradient cases as it is dominant in spectral analysis (not shown). Detail settings are in the section~\ref{sec:Eady-model}. After integrating the linear instability model for a sufficiently long time (200 day) from an arbitrary $k=5$ initial condition, the most unstable mode dominates all others.



The $\overline{u'v'}$, $\overline{u'\omega'}$ and $\overline{v'T'}$ induced by the most unstable modes are shown in the lower panels of Fig.~\ref{fig:eddy-momentum-transport}, Fig.~\ref{fig:eddy-vertical-momentum-transport} and Fig.~\ref{fig:eddy-heat-transport}. Clearly, the profiles of the eddy transports are similar to those simulated in the HS experiments. For the normal case, both types of model show a meridional momentum transport amplified around 200 mb (Fig.~\ref{fig:eddy-momentum-transport}a,c), an upward westerly momentum transport below 200 mb and a downward transport above (Fig.~\ref{fig:eddy-vertical-momentum-transport}a,c), and a heat transport peak near the surface and tropopause (Fig.~\ref{fig:eddy-heat-transport}a,c). However, for the reverse case, both models show bottom amplified meridional momentum transport (Fig.~\ref{fig:eddy-momentum-transport}b,d) and meridional heat transport (Fig.~\ref{fig:eddy-heat-transport}b,d), and a downward momentum transport below 200 mb (Fig.~\ref{fig:eddy-vertical-momentum-transport}b,d). Above 200 mb, eddies start to transport heat poleward (this is thermally-indirect), consistent with the weak residual cell in the upper atmosphere (Fig.~\ref{fig:residual-circulation-tropo}b). The streamfunctions of the most unstable modes are shown in the lower panels of Fig.~\ref{fig:eddy-structure}. Similar to the baroclinic eddy structure diagnosed from the HS experiments, the most unstable mode tilts westward (eastward) is upper (lower) troposphere amplified, in the normal (reverse) case.

The matching of the eddy momentum and heat transport profiles and the streamfunction structures is robust (not sensitive to the choices of wavenumber and latitude within 10 deg), indicating that the different eddy transport structures in HS experiments may stem from the difference in baroclinic modes, even without considering the wave propagation.
Actually, the top-amplified eddy structure is a general feature in our generalized Eady model when the wind shear and QGPV gradient have the same sign, and it holds even when setting beta and the wind shear to be constant across the domain, consistent with previous studies \cite[][]{Charney-1947:dynamics}. The bottom-amplified and eastward-tilted mode structure for the reverse case may be intuitively understood by turning the normal case Eady model upside down, but the vertical profile of the stratification turned out to be also important.
 
We think vertical structure of eddies is determined by the most unstable mode, different from AS15, who used WKB argument associated with wave propagation. The vertical scale of eddies is large compare to the troposphere depth (actually the vertical structure is almost barotropic throughout the troposphere), casting questions on accuracy of WKB approximation on which wave propagation theory is based. But ignoring this issue, one may view the two explanations as complementary views of the vertical eddy structure in the reversed case.

Like all other linear theories, the Eady model like setup assumes a small perturbation amplitude and a steady background state, which are not necessarily satisfied. In addition, the eddy structures diagnosed from the HS experiments do have some differences from the most unstable mode. The upper heat transport center in the Eady model normal case is stronger than the lower center, instead of the other way around; the $\overline{v'T'}$ becomes negative close to the upper boundary, which is not observed in the HS experiment; the eddy structure seems shifted upward for both cases. These may be related to the unrealistic solid-wall top boundary condition.

\paragraph*{Implications from the baroclinic eddy structure}

Based on the baroclinic eddy features investigated in the previous section, here we discuss their implications on the Ferrel cell, Hadley cell structure, and the surface wind pattern. We refer to the northern hemisphere unless otherwise noted.

\begin{figure*}[!tbh]
 \centering
 \includegraphics[page=2,width=\textwidth]{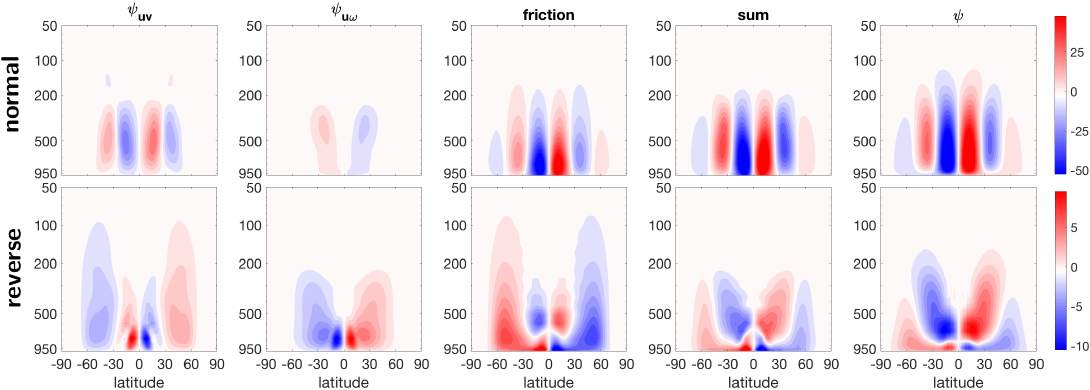}
 \caption{Meridional circulation decomposition into contributions by different processes. Shown are the dominant terms, from left to right are the meridional circulation associated with meridional eddy momentum transport $\overline{u'v'}$, vertical eddy momentum transport $\overline{u'\omega'}$, surface friction, and sum of these three terms, evaluated by solving Eq\eqref{eq:poisson}. The rightmost panels show the total meridional streamfunction from Eq\eqref{eq:total-psi}. Similarity of the right two panels indicates that the terms shown here explain the most of the total circulation.}
 \label{fig:decomposition-perpetual-annual}
\end{figure*}

\begin{figure*}[!tbh]
 \centering
 \includegraphics[page=3,width=0.6\textwidth]{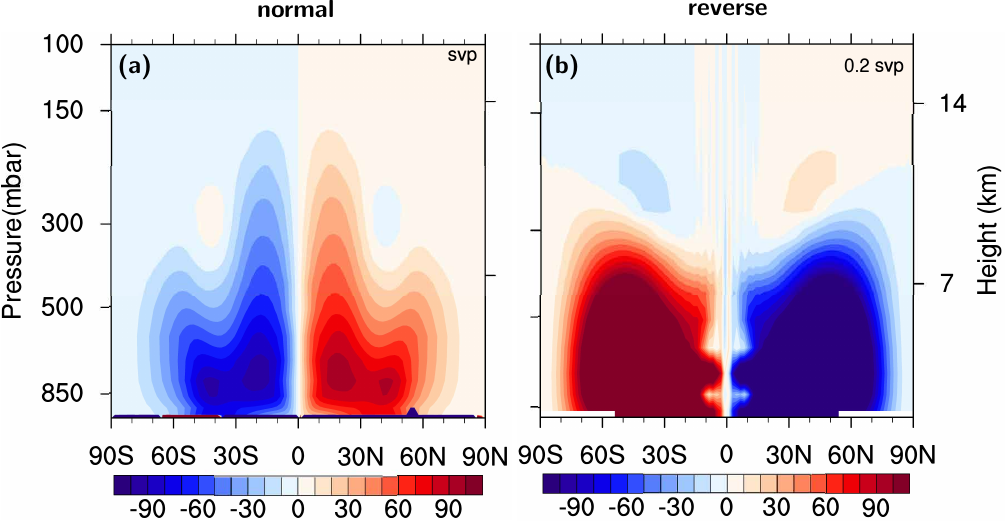}
 \caption{Same as Fig.~\ref{fig:meridional-circulation-tropo}(a,b), except for residual circulation.}
 \label{fig:residual-circulation-tropo}
\end{figure*}

\begin{figure*}[!tbh]
 \centering
 \includegraphics[page=4,width=0.6\textwidth]{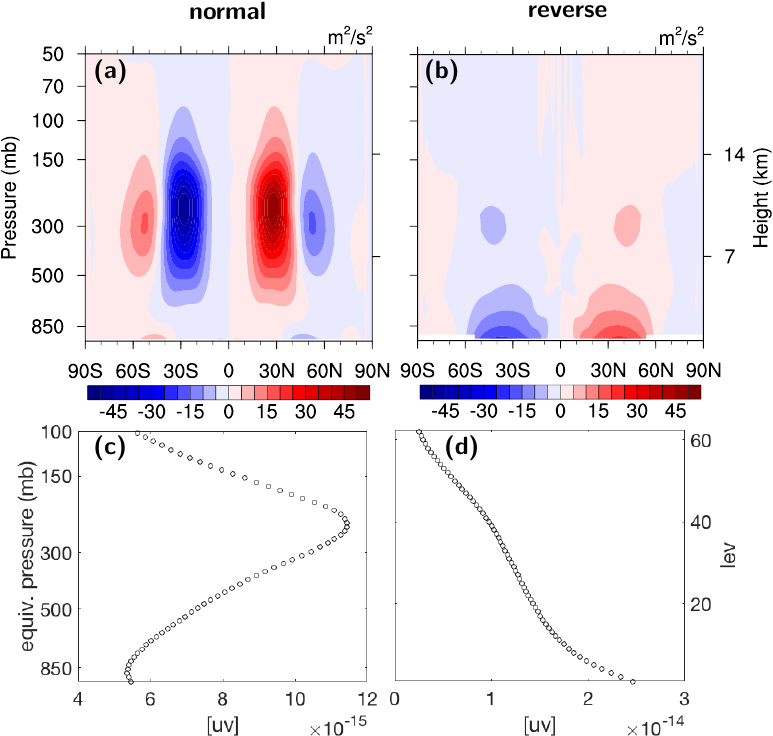}
 \caption{Upper panels (a,b) are meridional eddy transport of westerly momentum, $\overline{u'v'}$, where $\overline{\cdot}$ denotes zonal average, and prime denotes deviation from time mean zonal mean. Lower panels (c,d) show the $\overline{u'v'}$, in the generalized Eady model, under normal case and reverse case settings respectively. The sign of the profiles correspond to those at the south flank of wave source (45N and 55N for normal and reverse cases respectively).}
 \label{fig:eddy-momentum-transport}
\end{figure*}

\begin{figure*}[!tbh]
 \centering
 \includegraphics[page=5,width=0.6\textwidth]{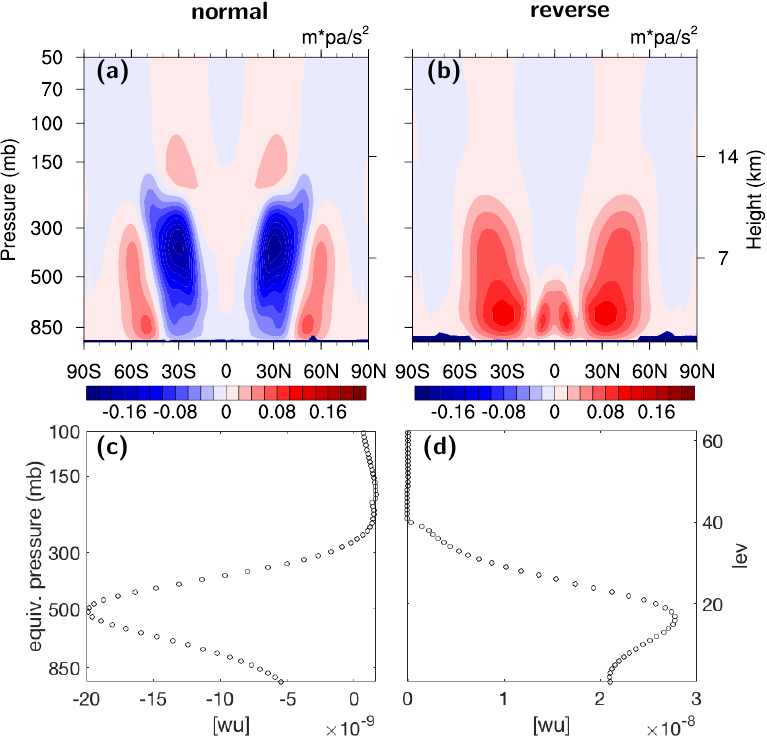}
 \caption{Same as Fig.~\ref{fig:eddy-momentum-transport} but for vertical eddy transport of westerly momentum, $\overline{u'\omega'}$.}
 \label{fig:eddy-vertical-momentum-transport}
\end{figure*}

\begin{figure*}[!tbh]
 \centering
 \includegraphics[page=6,width=0.6\textwidth]{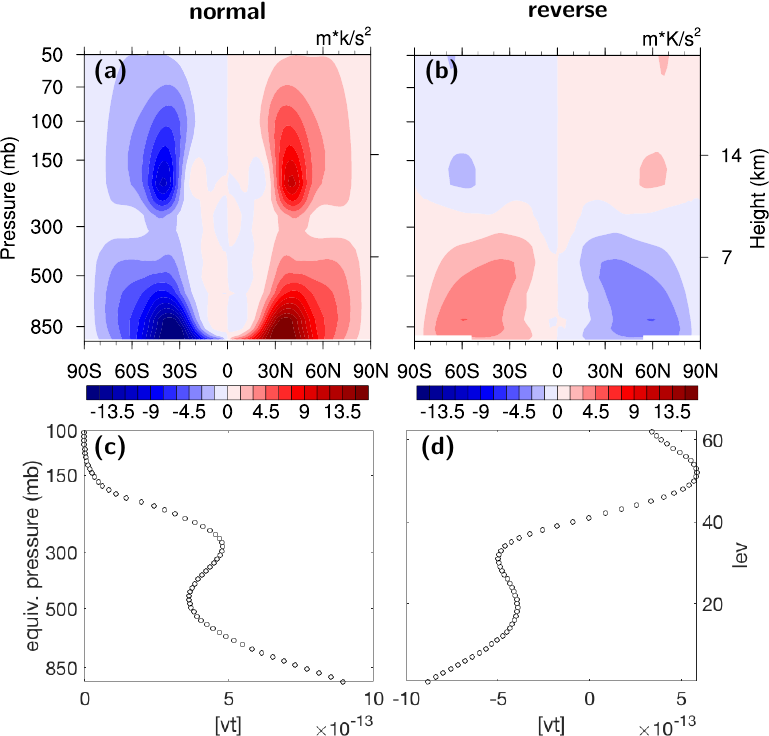}
 \caption{Same as Fig.~\ref{fig:eddy-momentum-transport}, but for meridional eddy heat transport, $\overline{v'T'}$. }
 \label{fig:eddy-heat-transport}
\end{figure*}

\begin{figure*}[!tbh]
 \centering
 \includegraphics[page=7,width=0.6\textwidth]{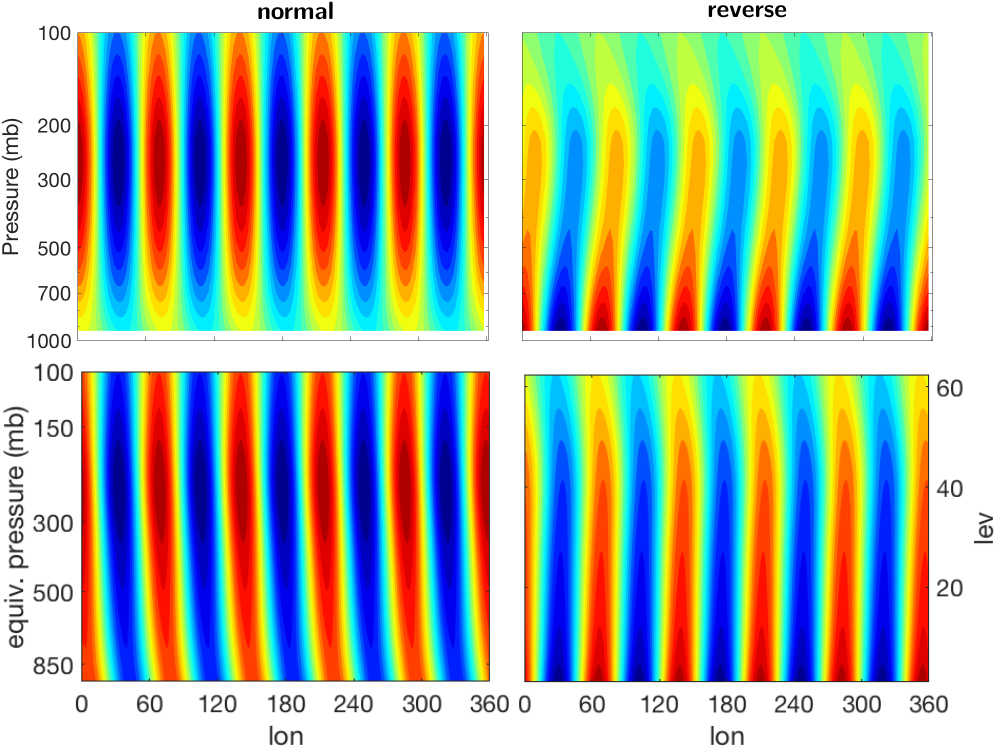}
 \caption{Barocinic Eddy structure for $k=5$ perturbation. Upper panels are diagnosed from Held-Suarez normal/reverse experiment, by projecting geopotential field at 45N/55N onto the timeseries of the geopotential at (45N/55N, 700 mb, 0E). Lower panels are the structure of the most unstable mode structure in the generalized Eady model, obtained by integrating the model for enough long time. Left panels are for the normal case and the right panels are for the reverse case. All plots are normalized to have unit maximum amplitude, thus the colorbar is omitted.}
 \label{fig:eddy-structure}
\end{figure*}

\begin{figure*}[!tbh]
 \centering
 \includegraphics[page=8,width=0.6\textwidth]{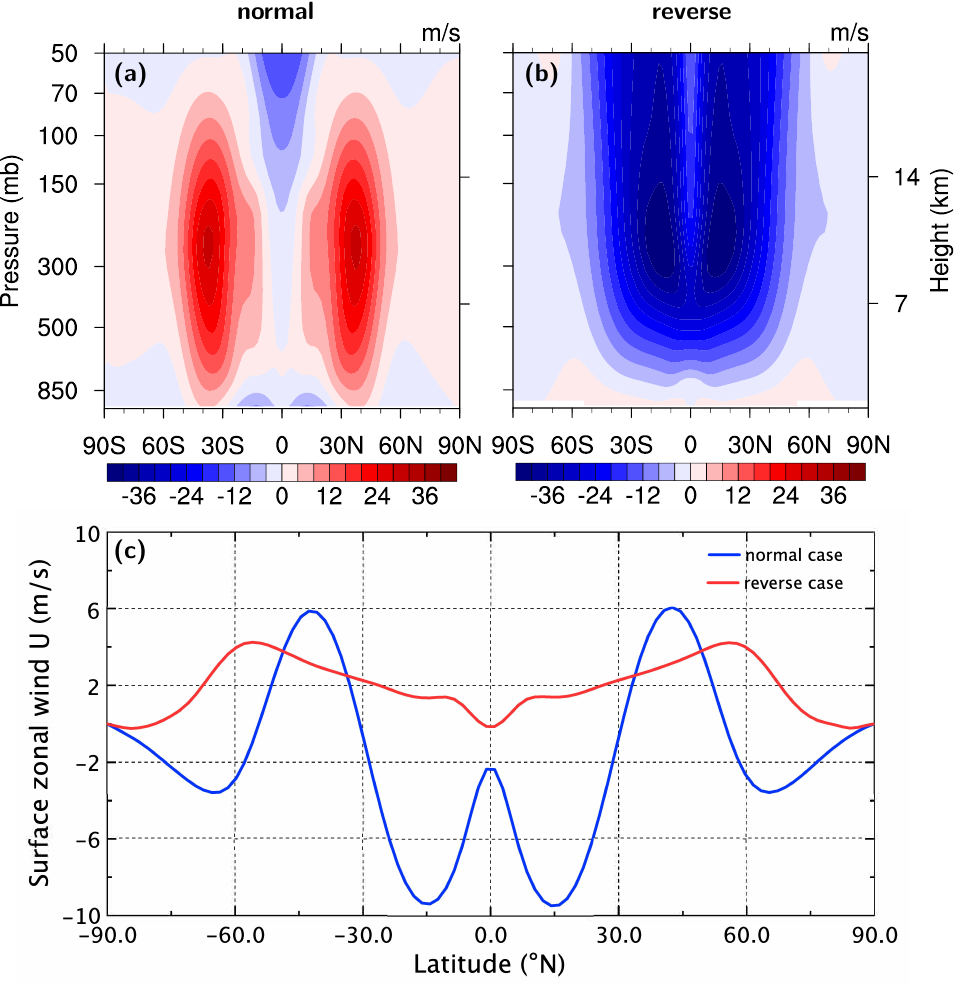}
 \caption{Climatological zonal mean zonal wind for (a) the normal case and (b) the reverse case, together with the surface wind latitudinal profile shown in (c).}
 \label{fig:U-climatology}
\end{figure*}

The surface wind in the reverse case is surprisingly similar to that in the normal case Fig.~\ref{fig:U-climatology}, with westerlies in the mid-latitudes and easterlies in the low and high latitudes. This surface wind pattern is required for the surface friction to balance the column-integrated eddy momentum convergence or divergence, outside the 30N/S where the momentum transport by the meridional overturning circulation is not dominant. Since baroclinic eddies tend to concentrate westerly momentum toward the mid-latitude wave source region, as long as the Coriolis coefficient is increasing with latitude \cite[the sign of meridional temperature gradient does not matter, e.g.,][]{Vallis-2006:atmospheric}, the surface winds in both cases show a easterly-westerly-easterly pattern from the equator to the pole. The surface westerlies in the reverse case, however, is much wider in latitude compared to that in the normal case, and this is related to latitudinal expansion of the baroclinic eddy activity as shown in Fig.~\ref{fig:eddy-heat-transport}. As shown in Fig.~\ref{fig:decomposition-perpetual-annual}, surface friction is the dominant driver for meridional circulation in both cases. Because the surface wind pattern is similar in the two cases, the associated meridional circulation patterns are also similar: clockwise-anticlockwise-clockwise from the Equator to the Arctic.

Ferrel cell (the meridional cell in mid-latitudes) is completely driven by friction in the reverse case and is driven by both $\overline{u'v'}$ and friction in the normal case. In the reverse case, surface friction takes westerly momentum from the surface air, drives it poleward, forming an anticlockwise circulation near the surface (Fig.~\ref{fig:decomposition-perpetual-annual}). For the normal case, in addition to the friction-driven circulation, baroclinic eddies concentrate westerly momentum toward the mid-latitude barolinic zone around 300 mb (Fig.~\ref{fig:eddy-momentum-transport}). Gaining momentum, the mid-latitude air is forced equatorward, also forming another anticlockwise circulation aloft. Such an anticlockwise circulation is thermally-indirect for a normal temperature gradient, but is thermally-direct for a reversed temperature gradient. That means, the Ferrel cell tends to restore the meridional temperature gradient in the normal case, while to reduce the gradient in the reverse case. As a result, the eddies in the reverse case do not need to be as strong to reduce the meridional temperature gradient to a similar level.

The Hadley cell (the meridional cell within 30N/S) can be explained in the similar manner. For both cases, friction play a dominant role. Eastward friction induced by surface easterly in the low latitudes drives the surface air parcel equatorward, forming a clockwise circulation. In the reverse case, a clockwise cell is thermally indirect, and it is added by the downward momentum transport by eddies. For the normal case, baroclinic eddies take westerly momentum away from the low-latitude upper air, contributing the rest 30\% of the total circulation. In the dry model we use here, contribution from diabatic heating is negligible, and this should not be expected to hold in a moist model especially when the latent heating is strong. Since the warmest region is in the low (high) latitudes in the normal (reverse) case, we expect the moist effect will affect more the normal case Hadley circulation. Very close to the surface and the equator, the surface pressure gradient and the surface friction dominantly balance each other. In the normal (reverse) case, the equator is the warmest (coldest) region, and thus forms an equatorial surface low (high) pressure center, pulling (pushing) air equatorward (poleward) and forming a thermally direct cell very close to the surface.
 
We also examine the sensitivity to the rotation rate by repeating the experiments under half and double the Earth rotation rate. The corresponding climatological patterns for zonal wind, eddy momentum transport $\overline{u'v'}$, eddy heat transport $\overline{v'T'}$, and meridional streamfunction are shown in Supplementary material Figs.~S2,S3. Although the number of jets and the width of Hadley cell change significantly with the rotation rate, the shallowness and weakness of eddy activity and Hadley cell in the reverse case still hold.

\subsection{Seasonal Cycle}
\label{sec:seasonal-cycle}

A high obliquity also causes a strong seasonal cycle. In this section, we therefore add a seasonal cycle to the Held-Suarez experiments, by allowing the radiative equilibrium temperature profile in Eq.\eqref{eq:Teq} to change with time. We find that during solstices, a much stronger thermally direct Hadley cell forms in both cases as is found by \cite{Linsenmeier-Pascale-Lucarini-2015:climate} and \cite{Ferreira-Marshall-OGorman-et-al-2014:climate}. But instead of being driven by eddy momentum transport as in the normal case and the reverse case under perpetual annual mean setup, it is driven by the eddy heat transport in the hemisphere with stronger temperature gradient. The annual mean Hadley cell under the configuration with seasonal cycle resembles that in the perpetual annual mean setup for both the normal and reverse cases, justifying the use of annual mean forcing in section~\ref{sec:Hadley-vanishment}. 

We design time-varying temperature profiles for both cases (normal and reverse), to just reverse the direction of meridional temperature gradient while keeping the amplitude unchanged. The purpose is to isolate the role played by the direction of temperature gradient separate from the amplitude, rather than to achieve a realistic representation. We set the center temperature to $T_c=315 (255)$ K as in the previous sections for the normal (reverse) gradient case. As the meridional variation of the static stability is negligible (section~\ref{sec:Hadley-vanishment}), we set the static stability coefficient $S=5$ K in both cases for simplicity. To reflect the meandering of the center latitude $\phi_c$ and the off-phase fluctuations of $T_y$ of the two hemispheres, we apply the following setup,
\begin{eqnarray}
  T_y^{NH}&=&T_y^{ann}(1+B~ \cos(2\pi~ t/P))^C\\
  T_y^{SH}&=&T_y^{ann}(1-B~ \cos(2\pi~ t/P))^C\\
  \phi_c&=&\phi_{c0}~ \cos(2\pi~ t/P),
\end{eqnarray}
where $T_y^{NH},\ T_y^{SH}$ are the meridional temperature gradient in the Northern and Southern Hemisphere respectively, $P=365$ day is the orbital period. $T_y^{ann}=-60$, $B=-0.9504$, $\phi_{c0}=-10$ for the normal case, and $T_y^{ann}=60$, $B=0.9504$, $\phi_{c0}=10$ for the reverse case. $C=0.2308$ for both. The maximum and minimum of $|T_y|$ are 70 K and 30 K respectively, and the maximum is achieved during winter (summer) solstice in the normal (reverse) gradient case. The center latitude shifts to the southern (northern) hemisphere in the normal (reverse) gradient case during boreal winter. To help visualize the above profiles, we show the seasonal variation of surface temperature in Fig.~S4 of the supplementary material, and show the temperature and potential temperature profiles at the boreal winter solstice in Fig.~S5. We used the same contour levels as in Fig.~4 of \cite{Ferreira-Marshall-OGorman-et-al-2014:climate}, who ran a couple ocean-atmosphere model with realistic radiation parameterization.

The mid-latitude eddies under the solstice forcing (not shown) are similar to that under the perpetual annual mean forcing, except that the magnitude of the eddies is stronger in the hemisphere with stronger meridional temperature gradient. Since the eddy activity is shallower and weaker in the reverse case (section~\ref{sec:baroclinic-mode}), the Hadley cell is also shallower and weaker (see also AS15).

\begin{figure*}[!tbh]
 \centering
 \includegraphics[page=9,width=0.6\textwidth]{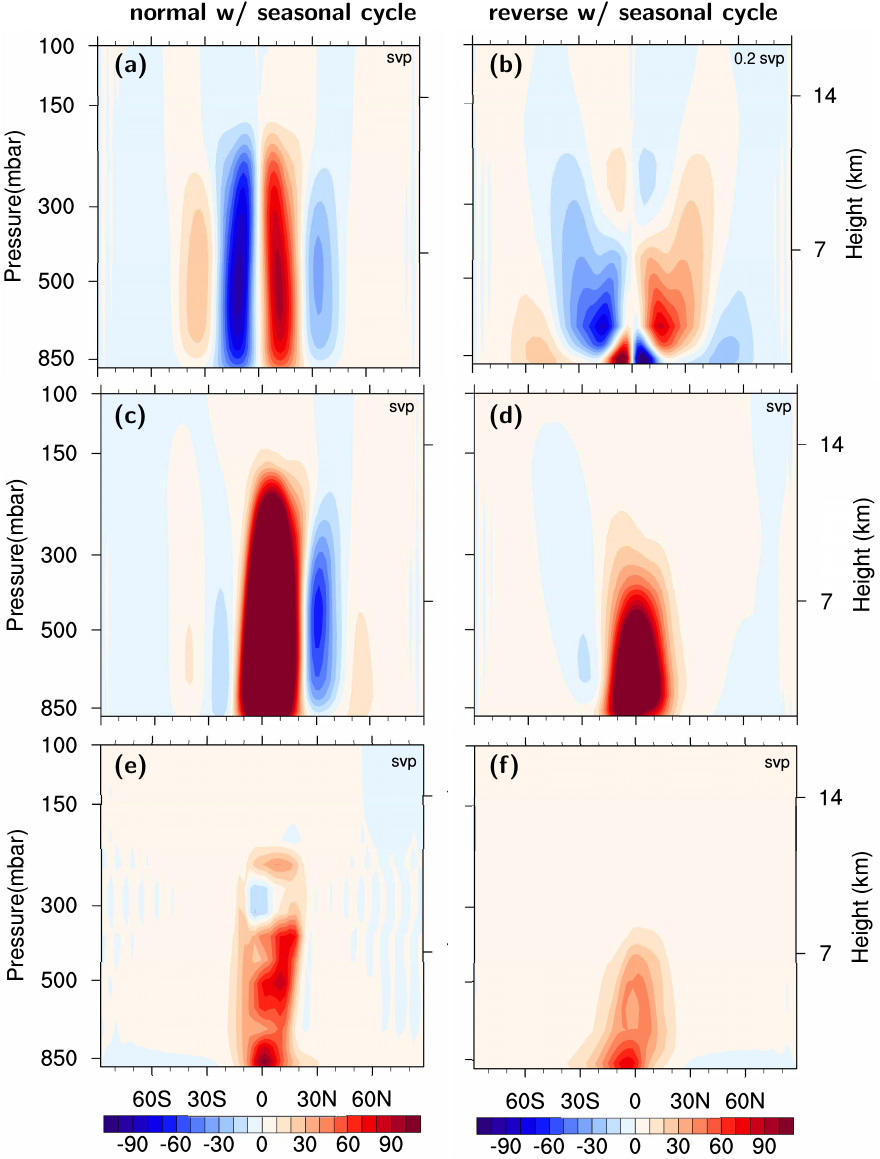}
 \caption{Zonal mean meridional circulation streamfunction for (a,c,e) the normal case and (b,d,f) the reverse case with a seasonal cycle. From top to bottom, shown are the annual mean, DJF climatology, and DJF climatology in the corresponding axisymmetric simulations. The streamfunction in panel (b) is in unit 0.2 svp, scaled up by 5 times to show the details. The maxima values are 95 svp, 25 svp, 405 svp, 212 svp, 104 svp, 79 svp, respectively for panel (a-f).}
 \label{fig:meridional-circulation-tropo-season}
\end{figure*}

The annual mean meridional circulation of simulations with seasonal cycle (Fig.~\ref{fig:meridional-circulation-tropo-season}a,b) turns out to be almost identical to those in the perpetual annual mean simulation (see Fig.~\ref{fig:meridional-circulation-tropo}a,b). Such similarity has also been observed for the Earth's atmosphere: the annual mean Hadley cell is almost identical to that at equinox \cite[][]{Levine-Schneider-Levine-2011:response, Walker-Schneider-2005:response}. The meridional circulation during DJF is shown in Fig.~\ref{fig:meridional-circulation-tropo-season}c,d. A stronger cross equator Hadley cell from the summer hemisphere to the winter hemisphere shows up in both the normal and reverse gradient cases, but compared to the normal gradient case, the Hadley cell in the reverse gradient case is shallower and weaker. The maximum of the meridional streamfunction in the normal and reverse gradient cases are 405 Svp and 212 Svp, respectively. The normal case Hadley cell is of the same magnitude as the observed Earth's Hadley cell during solstices, and the reverse case Hadley cell magnitude is consistent with the high obliquity simulation by \cite{Ferreira-Marshall-OGorman-et-al-2014:climate} where a much more sophisticated physics scheme was applied.

\cite{Linsenmeier-Pascale-Lucarini-2015:climate} also noticed a thermally-indirect annual-mean Hadley cell under 90 degree obliquity configuration, but it turned out to be an artifact of taking annual average. Their simulations also showed two very strong thermally-direct cells outside the thermally-indirect ones, which is much weaker in ours. First of all, our model is highly idealized, and thus care needs to be taken in the result interpretation. In addition, their model's ocean component is a slab ocean model, where the ocean heat transport (Q flux) is set to zero. As a result, the poleward heat transport needs to be accomplished by the atmosphere alone, and in fact, their solstice Hadley cell was two times stronger than in a fully coupled ocean atmosphere model \cite[][]{Ferreira-Marshall-OGorman-et-al-2014:climate}. 

\begin{figure*}[!tbh]
 \centering
 \includegraphics[page=10,width=\textwidth]{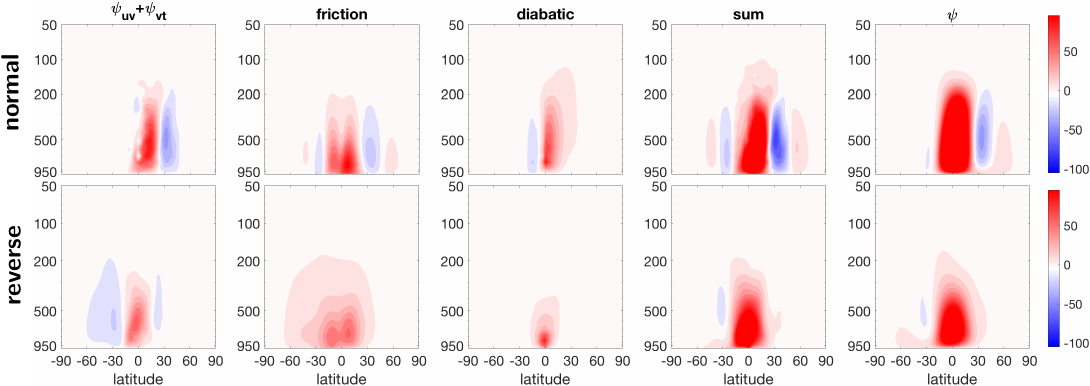}
 \caption{Similar to Fig.~\ref{fig:decomposition-perpetual-annual}, but for the solstice season in the seasonal varying experiments. Since the dominant processes driving the solstice circulation are different from those driving the annual mean. Shown from left to right are instead the meridional circulation associated with meridional eddy momentum and heat transport $\overline{u'v'}+\overline{v'T'}$, surface friction, diabatic heating and the sum of the three. The rightmost panels are the total meridional streamfunction from Eq\eqref{eq:total-psi}.}
 \label{fig:decomposition-solstice}
\end{figure*}

As before, we investigate the eddy effect through budget analysis and through the eddy-free simulations.
We solve Eq\eqref{eq:poisson} for the solstice season of both cases (see section.~\ref{sec:decompose-meridional-circulation} for details), and show the dominant terms, those associated with the sum of $\overline{u'v'}$ and $\overline{v'T'}$, friction, and Q, in Fig.~\ref{fig:decomposition-solstice}. Also shown are the sum of the above three dominant terms and the total streamfunction, the vertical integral of meridional mass transport. The matching of them indicates the above dominant terms explain most of the meridional circulation.

Diabatic heating contributes a significant part, but most of the meridional circulation is driven by friction and eddies. The patterns of $\overline{u'v'}$ and $\overline{v'T'}$ associated meridional circulation are similar, but $\overline{v'T'}$ contributes a greater portion (not shown). For the normal (reverse) case, the strongest eddy heat transport is located in the NH (SH), cooling (warming) the subtropics in the winter (summer) hemisphere. This helps restore the meridional temperature gradient in the low-latitudes and drive a thermally direct circulation there. Eddy momentum transport also helps. By taking westerly momentum away from the upper tropics and depositing it to the upper mid-latitudes, in the winter hemisphere of the normal case, it drives a thermally direct circulation in the tropics and the thermally indirect one in the winter hemisphere mid-latitudes. In the reverse case, westerly momentum transport happens in the summer hemisphere near the surface, and it also drives a thermally direct circulation.

To verify that the eddy heat transport is playing a dominant role, we also run the above experiments under an eddy-free configuration, and the meridional overturning circulation streamfunctions are shown in Fig.~\ref{fig:meridional-circulation-tropo-season}e,f. Compared to the full simulations, the streamfunction maximum (i.e., the total mass transport) drops by 75\% and 60\% in the normal and reverse case respectively, confirming the role of the eddy heat transport.

From the equinox to the solstice during the annual cycle, the Hadley cell evolves from being dominated by eddy momentum transport to being more dominated by eddy heat transport; in both circumstances, the eddy transport plays a dominant role. We note that the dominance of eddy effects could be an artifact of the highly simplified physics processes in HS model. Our results indicate that the tropical general circulation can be strongly affected by the extratropical dynamics where baroclinic eddies are generated, therefore considering the tropics and the mid-latitudes as a coupled system is necessary for a comprehensive understanding.
 
\begin{figure*}[!tbh]
 \centering
 \includegraphics[page=11,width=0.8\textwidth]{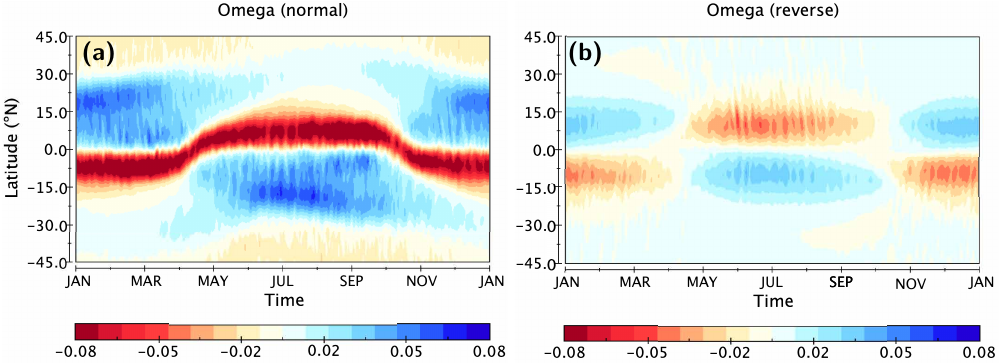}
 \caption{Seasonal cycle of the 500 mb vertical motion ($\omega$) for (a) normal case and (b) reverse case.}
 \label{fig:omegacyc-season}
\end{figure*}

We also show the seasonal cycle of the Inter Tropical Convergence Zone (ITCZ) represented by the corresponding vertical motions in Fig.~\ref{fig:omegacyc-season}. In the reverse case, the vertical motion is significantly weaker, and the latitudinal transition is more abrupt (Fig.~\ref{fig:meridional-circulation-tropo-season}b).

\section{Conclusions}
\label{sec:conclusions}

We investigated the general circulation of a atmosphere under a reversed temperature gradient (temperature increases toward the poles, reverse case), and compared it to that of an Earth-like atmosphere (normal case). The reverse temperature gradient setup is relevant to high obliquity planets with a surface that has high heat capacity, such as an ocean, as the temperature gradient from the equator to the pole was shown to be positive all year round \cite[][]{Jenkins-2001:high, Ferreira-Marshall-OGorman-et-al-2014:climate}. We found that the general circulation does not simply reverse when forced by a reversed meridional temperature gradient, and our objectives were to study the dynamics of eddies, and how eddy momentum and eddy heat transport explain the Hadley and Ferrel in this regime.

We employed an idealized \cite{Held-Suarez-1994:proposal} dry dynamic model, with the temperature being restored to an equilibrium temperature structure that increase or decrease toward the poles. The advantage of doing so is that it allows to isolate the role of the direction of the meridional temperature gradient by holding the temperature gradient magnitude and the stratification unchanged. 



We first compare the Hadley cell in the normal and reverse cases under a perpetual annual mean setup. Instead of being simply reversed, the Hadley cell in the reverse becomes thermally indirect and shallow, as oppose to a thermally direct and deep Hadley cell in the normal case. Its magnitude is over 4 times weaker than in the normal case, in spite of the amplitude of the thermal gradient being the same. The decomposition of the circulation into contributions by eddies, friction and diabatic heating suggests that both of the strong thermally direct Hadley cell in the normal gradient case and the weak thermally indirect Hadley cell in the reverse gradient case are driven mainly by the surface friction and by the poleward momentum transport by the mid-latitude eddies. Under eddy-free configuration, the asymmetry between normal and reverse cases' Hadley cell disappears. This suggests that eddie, via their momentum transport and their surface friction effects, play an important role in driving Hadley circulation. For the thermally indirect circulation in the reverse case, the low level circulation is driven mainly by surface friction, and most of the upper atmosphere meridional circulation is driven by meridional momentum transport $\overline{u'v'}$. Vertical eddy momentum transport $\overline{u'\omega'}$, which is small in the normal case, plays an important role in driving meridional circulation in the reverse gradient case. No matter in normal or in reverse case, the meridional circulation is mainly driven by momentum drag. These differences in meridional circulation finally gets down to the relative weakness and shallowness of the baroclinic eddies (and the eddy momentum transport) in the reverse case. 

The conclusion that eddy momentum transport concentrates near the tropopause in the normal case, while concentrated near the surface in the reverse case has been reached in AS15. AS15 suggests that the downward wave propagation in the reverse case, corresponding to the equatorward heat transport, leads to wave activity accumulation near the surface, where wave flux is redirected meridionally. In the normal case, poleward heat transport implies upward wave activity propagation and the accumulation of wave activities near the tropopause level explains the meridional momentum transport mainly takes place in the upper level. We here consider a complementary interpretation in which the tropospheric eddies are made of baroclinically unstable waves. The difference in the vertical structure of baroclinically unstable eddies under easterly shear (reverse case) and westerly shear (normal case) can account for eddies’ vertical structure including their vertical extent in the full GCM model. The vertical structure of baroclinically unstable waves is determined by the properties of the background flow, including profiles of stratification, zonal wind, and meridional QGPV gradient. Baroclinically unstable waves have to tilt against the vertical shear of the background zonal wind to tap the available potential energy embedded in the mean flow. Its amplitude and tilting angle are adjusted in a way to synchronize the Rossby waves in each layer and to guarantee each Rossby grow at the same rate. An idealized baroclinic model reveals that the lower (upper) troposphere part of the eddy perturbation is stronger in the reverse (normal) case. This explains that in the full GCM model the eddy heat and momentum transport are more upper (bottom) amplified in the normal (reverse) case. We further substantiate this interpretation by showing the eddy features (including the eddy momentum/heat transport and the vertical structure itself) diagnosed from the dry core experiments can mostly be captured by the most unstable mode in a generalized Eady model. 

The knowledge gained from the idealized baroclinic model help understand the following  features related to the asymmetry in the meridional circulation between the normal and reverse cases:

\begin{itemize}
\item Surface wind pattern determines the direction of surface friction, and the meridional circulation being driven. The fact that baroclinic instability is the source of mid-latitude eddies in both normal and reverse cases implies the accumulation of westerly momentum takes place in mid-latitude in both cases. Therefore, the surface wind pattern has to be easterly-westerly-easterly from the equator to the poles in both cases such that the excessive westerly momentum in mid-latitudes can be removed by surface friction processes which also help the atmosphere to gain westerly momentum outside baroclinic zone to balance the loss of westerly momentum there. Therefore, the surface friction helps form a meridional circulation pattern that is clockwise-anticlockwise-clockwise sequentially from the equator to the north pole.

\item The eddy driven Hadley circulation heads to opposite directions in normal and reverse cases, as the eddy momentum poleward transport occurs in the upper atmosphere and lower atmosphere respectively. It aids to the surface friction driven circulation in the normal case and cancels it in the reverse case upper atmosphere. This partially explained why the reverse case has a weak and shallower circulation compared to the normal case.

\item The meridional circulation in the reverse case is shallow also because the eddies are bottom amplified in the reverse case and thus there is no strong eddy drag in the upper atmosphere to drive circulation.

\item The resultant Ferrel cell in mid-latitudes is thermally indirect (direct) in the normal (reverse) case, acting to restore (reduce) the meridional temperature gradient there. This may partially explain why the eddy activity gets much weaker in the reverse case, which then leads to a much weaker meridional circulation.
\end{itemize}
 We also consider the strong seasonal cycle in a high obliquity climate. The annual mean meridional circulation with seasonal cycle resembles those without one, indicating that the perpetual annual mean setup may be relevant in understanding the annual mean climatology of a high obliquity planet. During the solstice, a much stronger thermally direct Hadley cell forms in both the normal and reverse cases, compared to the perpetual annual mean setup. Eddies again turned out to be important; but instead of through the eddy momentum transport as in the perpetual annual mean configuration, the solstice Hadley cells in the both cases were shown to be more driven by eddy heat transport. The strongest meridional temperature gradient is located in the summer (winter) hemisphere for the reverse (normal) gradient case, and the eddy heat transport in that hemisphere significantly enhances the hemispheric temperature contrast as imposed by the radiation, driving a strong cross-equatorial Hadley cell. When switching off the eddies, the solstice Hadley cells in the normal and reverse gradient cases are weakened by 75\% and 60\% respectively, indicating the dominant role played by eddies.

Our study uses a highly idealized model, with no explicit radiation scheme, no hydrological cycle, no ocean/ice dynamics, and no surface processes. While this allowed us to investigate the effects of the direction of the temperature gradient, isolated from the effects of its magnitude, including these components may change the general circulation. We also note that to maintain a reversed temperature gradient throughout the year, requires a very high obliquity (greater than 54$^\circ$), a surface with high heat capacity (which can be achieved by an ocean, or by low enough temperature) so that the radiative time scale is longer compared to the orbital period. An Earth-like planet with 70 or 90 degree obliquity was shown to satisfy such requirements \cite[][]{Jenkins-2001:high, Ferreira-Marshall-OGorman-et-al-2014:climate}. In addition, we intentionally only explore one single parameter, the sign of meridional temperature gradient, while holding other parameters to be the same (e.g., global mean temperature is Earth-like, the eccentricity effect is not considered). While this provides understanding of the effects of the temperature gradient, the results cannot be seen as a thorough investigation of the effects of high obliquity more generally.


\section{Acknowlegements}
\label{sec:acknowlegements}

This work was supported by the NSF P2C2 program, grant OCE-1602864, and by the Harvard Global Institute and Harvard Climate Solutions funds. ET thanks the Weizmann Institute for its hospitality during parts of this work. We would like to acknowledge high-performance computing support from Cheyenne provided by NCAR's Computational and Information Systems Laboratory, sponsored by the National Science Foundation.


%

\clearpage

 \bibliography{export}

\end{document}


\title{\Large{Supplementary Material for ``Tropical and Extratropical General Circulation with a Meridional Reversed Temperature Gradient as Expected in a High Obliquity Planet''}}
\date{}
\author{Wanying Kang, Ming Cai, Eli Tziperman}
\maketitle
 \renewcommand\thefigure{S\arabic{figure}}  

 \section{Vertical profiles used in generalized Eady model}
\label{sec:vertical-profile}
 
The analytical forms of zonal mean zonal wind ($U$), Brunt Vasala frequency ($N$), and meridional QGPV gradient ($Q_y$) are shown below:

For normal case,
\begin{align}
  \bar{U}&=28 \exp\left[-\frac{(z-0.57H)^2}{(0.5H)^2}\right] \label{eq:Uprof_norm}\\ 
    N&=\min\left[0.019-0.01~\sin{\left(\frac{\pi z}{0.65H}\right)},0.02\right] \label{eq:N2_norm}\\  
  Q_y&=\begin{cases}
    1.8e-10~\exp\left[-\left|\frac{z-0.5H}{0.16H}\right|^{1.3}\right], \hspace{1.25cm} \mathsf{z>0.2H}\\
    -2e-10+2.2e-10~\left(\frac{z}{0.2H}\right)^{0.2},\hspace{1.25cm} \mathsf{z\leq 0.2H}\\
  \end{cases}\label{eq:Qy_norm}
\end{align}

For reverse case,
\begin{align}
 \bar{U}&=max\{-14\sin\left(\frac{\pi z}{2.2H}\right) +3, -8\}\label{eq:Uprof_rev}\\
 N&=\min\left[0.015-0.008~\sin{\left(\frac{\pi z}{0.65H}\right)}, 0.022\right]\label{eq:N2_rev}\\
 Q_y&=\begin{cases}
    -0.7e-10~\exp\left[-\left|\frac{z-0.53H}{0.19H}\right|^{1.8}\right]+2e-11, \hspace{1.25cm} \mathsf{z>0.4H}\\
    -0.8e-10+1e-10~\left(\frac{z}{0.35H}\right)^{0.5},\hspace{1.25cm} \mathsf{z\leq 0.4H}
  \end{cases}\label{eq:Qy_rev}
\end{align}

\begin{figure*}
 \centering
 \includegraphics[page=12,width=\textwidth]{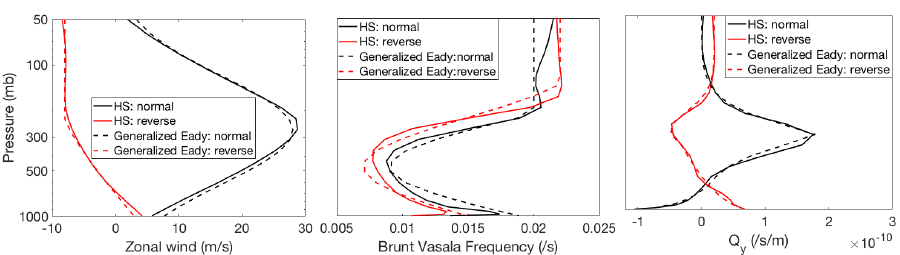}
 \caption{Comparison between Held-Suarez model output (solid lines) and the profiled used in generalized Eady model (dashed lines). Shown are vertical profile of (left) climatological zonal mean zonal wind, (middle) zonal mean Brunt Vasala Frequency, and (right) Quasi-geostrophic potential vorticity gradient profiles at 45N for normal case and 55N for reverse case. The normal cases are in black and the reversed cases in red.}
 \label{fig:U-sqrtN2-Qy-profiles}
\end{figure*}

These profiles are plotted and compared with Held-Suarez model output in Fig.~\ref{fig:U-sqrtN2-Qy-profiles}

\FloatBarrier
\section{Sensitivity to rotation rate}
\label{sec:sensitivity-rotation}

\begin{figure*}[!tbh]
 \centering
 \includegraphics[page=13,width=\textwidth]{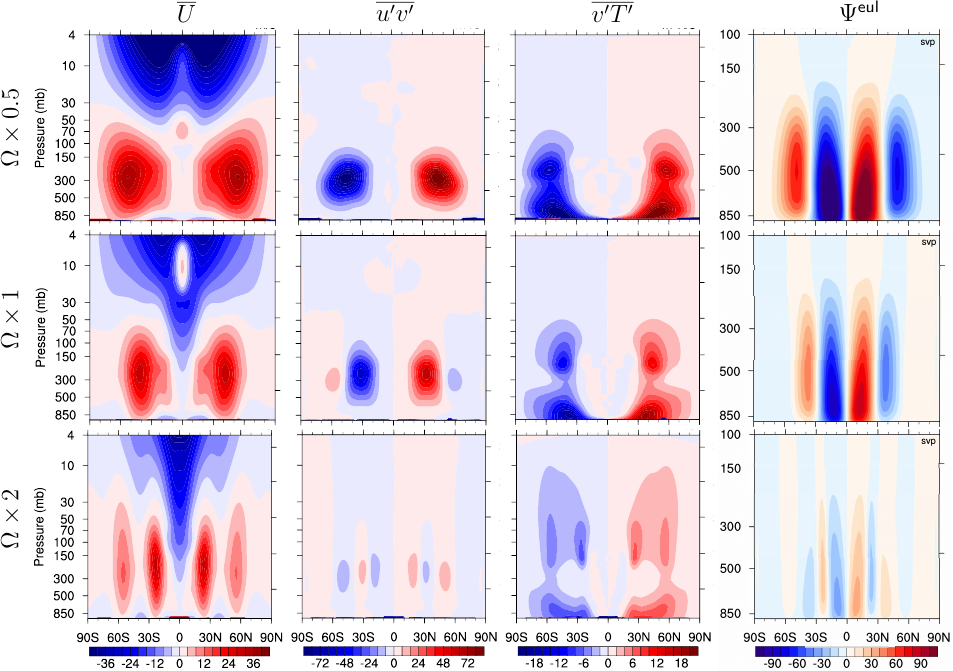}
 \caption{Climatology for normal case with different rotation rate. From left to right, shown are zonal mean zonal wind (in m/s), eddy momentum transport (in m$^2$/s$^2$), eddy heat transport (in K$~$m/s), and Eulerian meridional streamfunction (in svp). From top to bottom are results under half, unit, and double rotation rate of the Earth.}
 \label{fig:rotation-climatology-normal}
\end{figure*}

\begin{figure*}[!tbh]
 \centering
 \includegraphics[page=14,width=\textwidth]{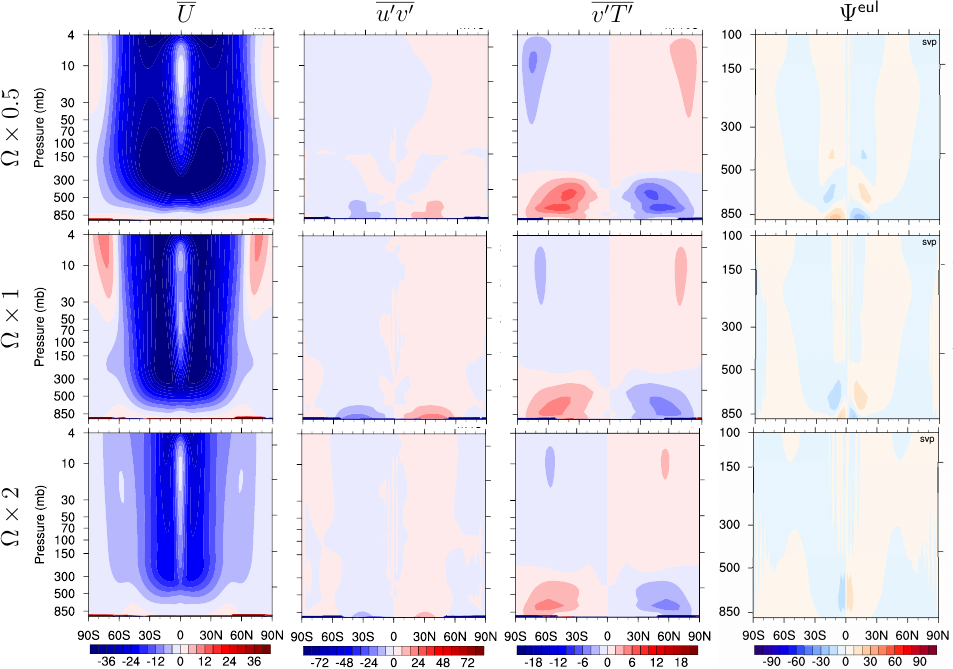}
 \caption{Same as Fig.~\ref{fig:rotation-climatology-normal}, but for the reverse gradient case.}
 \label{fig:rotation-climatology-reverse}
\end{figure*}

\FloatBarrier
\section{Seasonal varying forcing used in Held-Suarez experiments}
\label{sec:seasonal-forcing}
 \begin{figure*}[!tbh]
 \centering
 \includegraphics[page=15,width=0.7\textwidth]{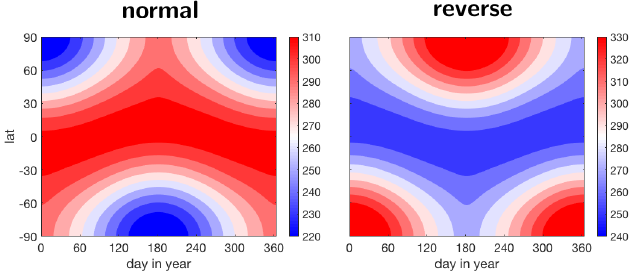}
 \caption{Seasonal variation of the radiative equilibrium temperature at surface, for normal and reverse cases.}
 \label{fig:Ts-season}
\end{figure*}

 \begin{figure*}[!tbh]
 \centering
 \includegraphics[page=16,width=0.7\textwidth]{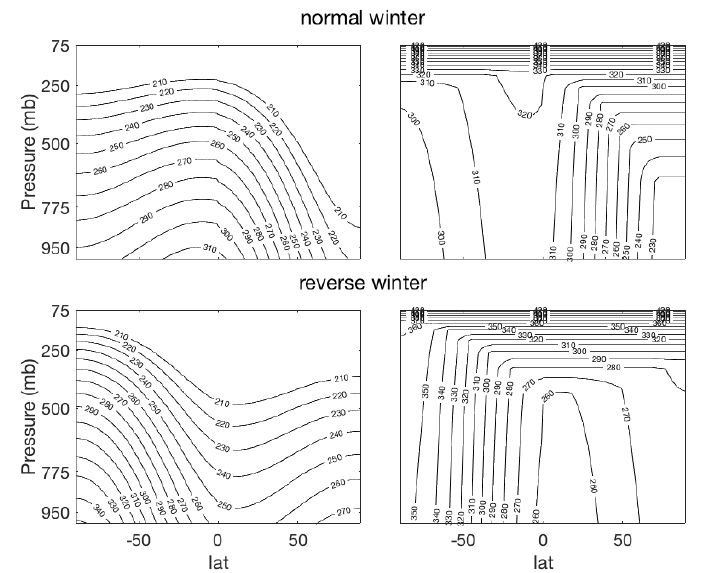}
 \caption{Radiative equilibrium temperature profile at northern hemisphere solstice for (upper) the normal case and (lower) the reverse case. Right panels the equilibrium potential temperature $\Theta_{eq}$ and left panels show the equilibrium temperature $T_{eq}$.}
 \label{fig:Teq-Theq}
\end{figure*}